\documentclass[preprint,showpacs,aps,prd,nofootinbib,floatfix,amsmath,amssymb]{revtex4-2}
\usepackage{graphicx}
\usepackage[utf8x]{inputenc}
\usepackage{color}
\usepackage{subfig}
\usepackage{multirow}
\usepackage{slashed}
\usepackage[dvipsnames]{xcolor}
\usepackage{siunitx}

\definecolor{brown}{rgb}{.7,.35,.1}
\definecolor{blueMBComm}{rgb}{0.1,0,0.5}

\begin{document}

\title{
	Effects of new physics on the Standard Model \\ parameters and vice versa
\footnote{Published in \textit{Tribute to Ruben Aldrovandi}, Editora Livraria da F\'\i sica, S\~ao Paulo, 2024; p. 57--566.}
}



\author{Mario W. Barela}%
\email{mario.barela@unesp.br
}
\affiliation{
	Instituto  de F\'\i sica Te\'orica--Universidade Estadual Paulista (UNESP)\\
	R. Dr. Bento Teobaldo Ferraz 271, Barra Funda\\ S\~ao Paulo - SP, 01140-070,
	Brazil
}

\author{V. Pleitez}%
\email{vpleitez@unesp.br, vicente@ift.unesp.br
}
\affiliation{
Instituto  de F\'\i sica Te\'orica--Universidade Estadual Paulista (UNESP)\\
R. Dr. Bento Teobaldo Ferraz 271, Barra Funda\\ S\~ao Paulo - SP, 01140-070,
Brazil
}

\date{10/12/2023
}
\begin{abstract}
This note discusses the matter of probing Beyond the Standard Model physics and how, to succeed in this quest, the interpretations of the Standard Model regarding observed phenomena must be utilized with caution. We give several specific examples of why this is necessary and assess general scenarios in which it is specially important. In particular, we call attention to the fact that once the Standard Model (SM) is finally replaced, the parameters of the new theory which embed those of the SM must be rederived directly from data instead of inherited through their expected relations with those of the SM.
\end{abstract}

\pacs{
}

\maketitle

\section{Introduction}
\label{sec:intro}

One could say that our understanding of the electroweak interactions among elementary particles evolved through around 117 years, beginning with the discovery of radioactivity (1986) and culminating in that of the Higgs boson (2012). A turning point in this process was the proposal of the formerly called ``Weinberg-Salam Model'' which was a two lepton families version of what it is now called ``eletroweak standard model''~\cite{Weinberg:1967tq,Salam:1968rm}. As is now known, however, this model was mathematically inconsistent: it featured the so-called triangle anomalies. Because it only possessed leptonic doublets and singlets (including massless neutrinos), it was incapable to cancel the anomaly factor, and, due to the chiral nature of electroweak interactions, could not upgrade the classical gauge symmetry to a quantum one.

From the theoretical point of view, the model could be considered mathematically consistent only when two generations of quarks were included~\cite{Glashow:1970gm} and the strong interaction described by the QCD~\cite{Rosner:1979hw} was properly formulated~\cite{Rosner:2002xi}. Experimentally, these advancements were confirmed when the quark charm $c$ and the asymptotic freedom of the strong interactions were discovered. In that newly formed model, the triangle anomalies cancel out generation by generation, thanks, in particular, to the color degrees of freedom. As an immediate accidental benefit of this development, the old question of why the muon must exist acquired a simple answer: because the model with only three quarks and two leptons is anomalous.

Note that, {\it at the time}, if the Weinberg-Salam model had been perceived to be anomalous, discarding the model in favour of some consistent alternative instead of predicting additional representation content would be, in principle, a viable position. The experimental observation which was, then, interpreted as the particles which were theorized in that context could have been found to correspond to different theoretical objects. Of course, with our current familiarity with the SM as an unimaginably accomplished fundamental theory, such remarks may sound odd or of little use, but, as we shall argue, it is an important notion to have in mind.

\section{Examples in the Standard Model construction}
\label{sec:first_examples}

For a further, less drastic thought experiment, consider the discovery of the $\tau$. The first evidence for the existence of a heavy lepton was the observation of the anomalous events~\cite{Perl:1975bf} 
\begin{equation}
e^++e^-\to e^++\mu^-+\textrm{misising energy}	
\label{mue}
\end{equation}
To interpret such events as intermediated by a sequential lepton one must impose a theoretical framework. In particular, it was assumed that the usual weak interaction is the ultimate description of nature, at least in the appropriate scenario. In practice, this amounts to considering that the hypothesized $\tau$ decays are mediated only by the $W$ vector boson alone, {\it i.e.}, that the lepton-$W$ and quark-$W$ vertices have the standard electroweak couplings in the form of the four-fermion effective interactions -- and that the physical process may be subsumed to this sole contribution. Decays of the $\tau$ through the electromagnetic interaction such as $\tau\to e(\mu)+\gamma$ or through exotic interactions, for instance, are forbidden and irrelevant, respectively. There are additional non general inputs needed for the prediction of the $\tau$ decay modes: see, for instance Ref.~\cite{Perl:1976rz}. The process in Eq.~(\ref{mue}) may then be interpreted as
\begin{equation}
\begin{gathered}
e^++e^-\to \gamma^*\to \tau^++\tau^-, \\
\tau^+\to e^++\nu_e+\bar{\nu}_\tau, \qquad \tau^-\to \mu^-+\bar{\nu}_\mu+\nu_\tau,
\label{mue2}
\end{gathered}
\end{equation}
being $\tau$ a sequential spin-1/2 lepton. The possibility of spin-0 particles was ruled out because the rates of $\tau\to e^-+X$ and $\tau\to \mu+X$, where $X$ are hadrons, are equal within experimental error.
Additionally, the $e^{+} {} + {} e^{-} \to W^+ + W^-$ channel was ruled out by the energy distribution of the electron and muon~\cite{Perl:1976rz,Perl:1979pb}. To synthesize, although the process $e^++e^-\to \tau^++\tau^-$ relies only on precise QED calculations, the $\tau$ cannot be detected as a long-lived particle, but must instead be reconstructed from its final-state products which involve weak interactions and undetectable neutrinos produced only by the $W$ vector boson~\cite{Zani:2023ngd,Belle-II:2023izd}.

Indeed, these remarks could be made regarding any particle of the latest generations of the SM or the massive vector bosons. The $W^\pm$, for instance, was discovered in 1983 through identification of its leptonic decay channels: $W^\pm \to  e^\pm \nu_e(\bar{\nu}_e)$ in both UA1 and U2 and  $W^\pm \to  \mu^\pm \nu_\mu(\bar{\nu}_\mu)$ in UA2~\cite{DiLella:2015yit}. In the theory that replaces the SM, or New Standard Model (NSM), however, a particle that is identifiable with the $W^\pm$ could have exotic decay channels.

The lesson we want to emphasize is that high energy physics is model-loaded.
This has not always been necessarily the case. For instance, the masses of the electron and muon were measured isolating the electromagnetic theory, quite generally, and no specific non-Abelian model was assumed. In particular, note that the electromagnetic interaction is the only force to posses a fully classic counterpart. This was not the case of the $\tau,c,b,t,W^\pm,Z^0,H^0$ particles whose masses were interpreted in the context of the standard model. To strengthen the intuition behind these philosophy, let us put forward a few more examples.

A given model is determined by its gauge, accidental symmetries and representation content. This last element implies that the electroweak standard model restricted to two matter generations (2ESM) is, by construction, different from the current three matter generations theory (3ESM). With two generations, the quark sector possesses a single measure of mixture, the so-called Cabibbo angle $\theta_\mathrm{c}$~\cite{Cabibbo:1963yz}, related to the unitary mixing matrix by 
\begin{equation}
\vert V_{ud}\vert^2+\vert V_{us}\vert^2 \equiv \cos^2 \theta_\mathrm{c} + \sin^2 \theta_\mathrm{c} = 1.
\label{cabibbo}
\end{equation}
In the context of the 3ESM, however, the mixtures are measured by three angles and one physical phase. In this case, the analogous of Eq.~(\ref{cabibbo}) is
\begin{equation}
\vert V_{ud}\vert^2+\vert V_{us}\vert^2+\vert V_{ub}\vert^2=1,
\label{cabibbo2}
\end{equation}
where, as determined phenomenologically,
\begin{equation}
	\vert V_{ud}\vert^2+\vert V_{us}\vert^2\lesssim 1, \quad \vert V_{ub}\vert^2\ll 1.
	\label{cabibbo3}
\end{equation}
We see that interpreting phenomenology within a different model induce a small but not trivial change in the new analogous parameters. If we had simply embedded the 2ESM into the 3ESM hypothesis, we would obtain $V_{ub}=V_{cb}=V_{tb}=0$ (and the same for the remaining top parameters). Of course, the new model also introduces a new parameter (a physical phase) which induces strictly new physics ($CP$ violation!). Note, in particular, that the parameter analogous to the Cabibo angle in the new model is altered.

At this point, one could argue that this is all obvious or innocuous. Indeed, a model proposed to replace the SM should reproduce its good predictions and, in one way or another, imply it as an approximation at energy scales of up to around $\SI{1}{TeV}$. The theoretical mechanism through which this is usually built-in new models is that of deeming exotic particles to be heavy (after all, the SM accommodates the current measured cross sections without the need for extra particles), and concluding that their effects are completely decoupled~\cite{Appelquist:1974tg}, or they are weakly interacting. As final consequence, then, the parameters (specially effective ones) in the new model which are derived from parameters of the SM may be evaluated using their known values. Although this should be a fair assumption in a large sector of theory space, it is not a general fact (to start, the decoupling theorem may not guarantee the insignificance of exotic effects in every situation~\cite{Toussaint:1978zm,Arco:2023sac}, but this is not the only flaw in the argument above). 

Another example, of a parameter that should certainly change if new singly charged scalar and/or vector bosons are discovered and appended to the SM context is the mass of the top quark. This would provide a possible explanation as to why its mass seems exaggeratedly large in relation to the other particles masses in the model. Moreover, this would represent additional evidence (which has resisted over 20 years) for the discrepancy of $2.4\sigma$ in the muon $g-2$ factor~\cite{Muong-2:2023cdq}.

\section{Patterns to avoid in the Beyond the Standard Model phenomenology}
\label{sec:upp}

Ultimately, our discussion concerns theoretical occurrences which may cause phenomenological predictions to not apply to the models under consideration. Let us analyse three general patterns of the appearance of this phenomenon. For that, consider the usual interpretation of experimental measurements in the context of exploratory phenomenology, which naturally also applies to the examples already given. In such scenarios, there exists, for instance, a statistically significant upper limit on the cross section of some convenient process. In the ideal scenario, this process isolates an exotic contribution with little competing background. The observation is then cast in the form of a prediction resulting from a minimal effective model, and the upper limit on the cross section is translated into a limit on new physics parameters, such as masses or mixing angles. In this stage, two steps (which are ultimately equivalent) are tacitly implied by the usual procedure: choice of effective model and simplification. As a first example consider the search for a doubly-charged vector boson. The preferred channels for the observability of its effects, in general, involve its interactions with charged leptons. Its form is often assumed to be equivalent to~\cite{Corcella:2018eib,Corcella:2017dns,RamirezBarreto:2011av}

\begin{equation}\label{eq:Uint0}
\mathcal{L}_{U\ell\ell}= \sum_{a} g_{U\ell\ell}\, \bar{\ell_a^c}\gamma^\mu P_L \ell_a\, U_\mu^{++} + \mathrm{H.c.},
\end{equation}
where $P_L$ is the left-handed projector, $\ell_a$ is a lepton field ($a = e,\mu,\tau$) and $g_{U\ell\ell}$ is an effectively arbitrary perturbative coupling. The interaction above, however, is far from skeptical or model independent. Indeed, the most general relevant interaction allowed by Lorentz invariance, renormalizability and reality of the action is given by~\cite{Barela:2022sbb}


\begin{eqnarray}
\mathcal{L}_{U\ell\ell}&=&\sum_{b> a} g_{U\ell\ell} \left\{\bar{\ell_a^c}\gamma^\mu [P_L (V_U)_{ab}-P_R (V_U)_{ba}] \ell_b\thinspace U_\mu^{++} + \bar{\ell_a}\gamma^\mu [P_L (V_U^\dagger)_{ab}\right.\nonumber \\&-&\left.P_R (V_U^\dagger)_{ba}] \ell_b^c\thinspace U_\mu^{-\,-}\right\}+ \sum_{a} g_{U\ell\ell} \left\{\bar{\ell_a^c}\gamma^\mu [P_L (V_U)_{aa}] \ell_a\thinspace U_\mu^{++} \right.\nonumber \\&+&\left.\bar{\ell_a}\gamma^\mu [P_L (V_U^\dagger)_{aa}] \ell_a^c\thinspace U_\mu^{-\,-}\right\}
\end{eqnarray}

The major difference between this form and that of Eq.~(\ref{eq:Uint0}) is the unitary mixing matrix $V_U$, which, by naturalness principles, cannot be ignored. Moreover, $V_U = 1$ could only be the case for the simplest mass matrices -- diagonalizable by orthogonal transformations instead of by biunitary ones -- which exclude popular $U^{\pm \pm}$ models. 

Notice that, even if the searched reaction involves diagonal interactions, such as the case of $pp \to e^+e^+e^-e^-$ or $pp \to \mu^+ \mu^+e^-e^-$, and the $g_{U\ell\ell}$ coupling is left free, there is still an imprecision as the $U\ell\ell$ interaction is not universal and may couple the boson to different flavors with different strength. What we have underlined is that a popular framework for the interpretation of experimental results in the search for the $U^{\pm\pm}$ {\it definitely} produces inaccurate predictions for the exact parameters of some underlying theory. Of course, such simplifications may represent an acceptable approximation and translate to true, useful constraints, but the risk that it causes harmful distortions is also real. 

The second pattern we mentioned may be seen in studies which do not scan over $g_{U\ell\ell}$ and, at least implicitly, assume the 3-3-1 model. Comparing Eq.~(\ref{eq:Uint}) with the prediction of the 3-3-1, one obtains $g_{U\ell\ell} = g_{3L}/\sqrt{2}$, where $g_{3L}$ is the $SU(3)_L$ coupling. This is then used as a benchmark, equating $g_{3L}$ to the standard model $SU(2)_L$ coupling $g^{\mathrm{SM}}_{2L}$. This may source theoretical errors by several mechanisms. To start, the 3-3-1 $g_{3L}$ is expected to be equal to $g_{2L}$ at a single matching scale~\cite{Barela:2023oyp}, not identically. At the few TeV energies where the relevant process is explored, $g_{3L}$ should be used directly. Its value must be obtained from its running and matching with $g_{2L}$ (this process should involve extra unknown parameters). Additionally, the 3-3-1 coupling $g_{2L}$ is not generically equal to the SM $g^{\mathrm{SM}}_{2L}$ at any arbitrary scale, as their running may involve distinct group factors and contributing degrees of freedom. 

Once more, the simplification of setting $g_{U\ell\ell} = g^{\mathrm{SM}}_{2L}/\sqrt{2}$ may be a good approximation in the LHC era. But what about neutral current or even purely electromagnetic processes at high scales, which, within the 3-3-1 context, involve $g_X$? This coupling should be related to $g_Y$ carefully (which is rarely accomplished) and, differently from $g_{3L}$, runs aggressively with energy. As a result, failing to interpret experimental input within the framework one wishes to constrain in a precise manner may again lead to unreliable conclusions. 

Finally, the last pattern is even more general and applies to cases discussed before. For the sole sake of simplicity, a new experimental result is usually translated into a requirement on some parameter space through modelling with a single exotic degree of freedom. This relies on the hypothesis of absolute dominance (within expected precision) of the chosen exotic particle with relation to the contribution of other exotic concepts. This is, however, not general and ultimately amounts to a choice of model in which additional particles are absent or have effects which are pushed to sufficiently higher regimes. This usually represents a reasonable benchmark choice and is rarely general, so that, when possible, the extra contributions of a specific model should be assessed simultaneously.

The presence of subdominant, not accounted for, effects, however, can not only strengthen the constraints on masses, for instance, as in the case of the top we discussed -- it can also {\it weaken} it. Indeed, destructive interference effects which overcome the pure contribution of the second particle are possible and allow for smaller masses to become allowed, in principle. It has been explicitly verified, in a model independent analysis over an exotic sector with a doubly charged vector bilepton, a flavor changing neutral scalar and a doubly charged one, that interference with the neutral scalar may weaken the bound on the $U^{\pm \pm}$ by up to 20\%~\cite{Barela:2022sbb}. This is another very general way in which disregarding the most complete form of some specific framework may render predictions less useful.

In summary: the ideal phenomenological program, in the sense of the reliability of its results, is one which is framework specific. This is, of course, not possible in every situation, as some theoretical concepts are ubiquitous or, on the contrary, are not known to exist within a specific model. Additionally, to examine some hypothesis in a model independent fashion is not only useful but, sometimes, the obvious preferred choice, as it is impractical to reevaluate each established concept in every analysis. However, the lesson of our discussion is that once some model is actively proven to be a viable replacement of the SM or is under focused consideration, every phenomenological statement should be made within its own context. This means not only that an experimental input must be interpreted through complete parametrization of its physics, but also that values for old parameters (such as the SM $g_{2L},g_Y$ in the example above) {\it should not} be carried to the NSM. Instead, the analogous, dependent or parameters which embed the old ones should be rederived by a new investigation of the original phenomenological information.

\section{Examples in classical physics}
\label{sec:oldp}

In fact, the issues discussed in Sec.~\ref{sec:first_examples}  are not exclusive to quantum field theories. For instance, one could realize, in the context of classical mechanics, that some process is not well described by the conservation of the usual kinetic energy $T=mv^2/2$. One could then go one step further and propose the simple polynomial extension $T=mv^2/2+\kappa mv^4$ as a new dynamical model, which we could call Refined Newtonian. One would test this against observations and fit the new $\kappa$-parameter. Now, suppose, in the thought experiment, that Special Relativity was eventually discovered and found to be a better theory. The $\kappa$ of Refined Newtonian dynamics is naturally embedded in Special Relativity as a low order parameter in the expansion of $\gamma$ of $T=mc^2(\gamma-1)$ in powers of $v/c$. However, $\kappa=3/8c^2$ is now a calculable factor and its value differs from the one found within the framework of Refined Newtonian theory.
 
Another example in older physics is the Fresnel drag coefficient, $f=1-1/n^2$, where $n$ is the refraction index of the substance, proposed to explain the negative result of the Arago experiment, the first attempt at measuring the absolute motion of the Earth. According to Fresnel, the ether inside a substance is partly dragged when the substance moves with respect to the exterior ether. If a transparent substance moves with ‘absolute' velocity $v$, the ether within it moves with absolute velocity $v_\mathrm{drag}=fv$. This was, at the time, a satisfactory (phenomenological) solution to the conclusion of the Arago experiment. In some sense, this hypothesis could be considered as a `Refined' Maxwell theory, an extension of it. Only with the appearance of the alternative theory, the theory of special relativity, it was possible to understand the hypothesis of Fresnel~\cite{peter}. Fresnel's $f$ coefficient appears just as an approximation when $v\ll c$.

Another example of the case studied in Sec.~III is the precession of Mercury orbit. At the end of XIX century, for explaining that anomaly, there was some proposals which assumed modifications of the Newton Gravitation Law which would depend on velocity, but the results gave always a value for this anomaly below the observed one. The theory of general relativity (TGR) also provides for such a modification of the Newton law depending on the velocity,  but it gives the correct result. That is, it was not enough to propose such a potential depending on the velocity. There was only one, that of the TGR.

\section{Conclusions}
\label{sec:con}

After what was argued above, it would not be surprising if the W boson had a mass that differs from the one obtained in the context of the SM~\cite{CDF:2022hxs}, or the Cabibbo angle turns out to be different from the accepted value. The first case could be an evidence in favour of models such as the left-right symmetric ones~\cite{Diaz:2020pwh} or of some supersymmetric theories~\cite{Rodriguez:2022wix}. The second disparity, of the Cabibbo anomaly, in turn, could be explained by the existence of new particles, such as, for instance, leptoquarks~\cite{Kirk:2023oez}. 

In conclusion, although simplifications -- such as single exotic particle lagrangians with no mixing -- are understandable and necessary in Beyond the Standard Model physics, we urge phenomenologists to make an effort to precisely define a framework in order to derive and state results. Furthermore, once a viable replacement to the SM is finally confirmed, it should not be seen as an ```extension" of it, but as an alternative model altogether, which includes the SM as an approximation. Many of the parameters that were interpreted and fitted in the context of the SM should then be reconsidered. Which model, if any, would be that?

\section*{Acknowledgments}

MB is grateful to CNPq for the financial support.

\end{document}